\def\year{2017}\relax
\documentclass[letterpaper]{article} 
\usepackage{aaai17}  
\usepackage{times}  
\usepackage{helvet}  
\usepackage{courier}  
\usepackage{url}  
\usepackage{graphicx}  
\usepackage{relsize}
\usepackage[latin1]{inputenc}
\usepackage[english]{babel}

\usepackage{graphicx}
\usepackage[belowskip=0pt,aboveskip=0pt]{subcaption}
\usepackage{amsmath, amsthm, amssymb}
\usepackage{textcomp}
\usepackage{stmaryrd}
\usepackage{upgreek}
\usepackage{bm}
\usepackage{cases}
\usepackage{mathtools}
\usepackage{arydshln}
\usepackage{multirow}

\usepackage[pagebackref=true,breaklinks=true,letterpaper=true,colorlinks,bookmarks=false,citecolor=purple]{hyperref}

\usepackage{algorithm}
\usepackage{algpseudocode}

\usepackage[normalem]{ulem}

\usepackage{tabularx}
\usepackage{multirow}
\usepackage{rotating}
\usepackage{booktabs}

\usepackage{enumitem}
\usepackage[olditem,oldenum]{paralist}

\usepackage{alltt}
\usepackage{listings}

\usepackage{mysymbols}

\usepackage{url}
\usepackage{xspace}
\usepackage{comment}
\usepackage{afterpage}
\usepackage{pdfpages}
\usepackage{framed}
\usepackage{fancybox}
\usepackage{cuted}

\newcommand{\xhdr}[1]{\par \noindent
\textbf{#1}
}
\newcommand{\Qbot}{\textsc{Qbot}\xspace}
\newcommand{\Abot}{\textsc{Abot}\xspace}

\newcommand{\A}{\textsc{Alice}\xspace}

\newcommand{\Asl}{\textsc{Alice}$_{\texttt{SL}}$\xspace}

\newcommand{\Arl}{\textsc{Alice}$_{\texttt{RL}}$\xspace}
\newcommand{\GW}{GuessWhich\xspace}

\newcommand{\gw}{GuessWhich\xspace}
\newcommand{\hum}{\textsc{H}\xspace}

\newcommand{\reffig}[1]{Fig.~\ref{#1}}

\begin{document}

\title{Evaluating Visual Conversational Agents via Cooperative Human-AI Games}
\author{
Prithvijit Chattopadhyay$^1$\thanks{Work done at Virginia Tech.}\thanks{The first two authors (PC, DY) contributed equally.} \,\,
Deshraj Yadav$^1$\footnotemark[1]\footnotemark[2] \,\,
Viraj Prabhu$^1$\footnotemark[1] \,\,
Arjun Chandrasekaran$^1$ \,\,\\
\fontsize{12}{14}\selectfont \textbf{Abhishek Das}$^1$ \,\,
\textbf{Stefan Lee}$^1$\footnotemark[1] \,\,
\textbf{Dhruv Batra}$^{2,1}$ \,\,
\textbf{Devi Parikh}$^{2,1}$ \\ \\
\fontsize{12}{14}\selectfont $^1$Georgia Institute of Technology \quad
$^2$Facebook AI Research \\
{\tt\footnotesize \{prithvijit3, deshraj, parikh\}@gatech.edu} \\
\tt\footnotesize \href{http://visualdialog.org}{visualdialog.org}
}

\maketitle
\begin{abstract}
As AI continues to advance, human-AI teams are inevitable.
However, progress in AI is routinely measured in isolation, without a human in the loop.
It is crucial to benchmark progress in AI, not just in isolation, but also in terms of how it translates to helping humans perform certain tasks, \ie, the performance of human-AI teams.

In this work, we design
a cooperative game -- \GW~-- to measure human-AI team performance in the specific context of the AI being a visual conversational agent. \GW involves live interaction between the human and the AI. The AI, which we call \A, is provided an image which is unseen by the human. Following a brief description of the image, the human questions \A about this secret image to identify it from a fixed pool of images.

We measure performance of the human-\A team by
the number of guesses it takes the human to correctly identify the secret image after a fixed number of dialog rounds with \A.
We compare performance of the human-\A teams for two versions of \A. Our human studies suggest a counter-intuitive trend -- that while AI literature shows that one version 
outperforms the other 
when paired with an AI questioner bot, we find that this improvement in AI-AI performance does not translate to improved human-AI performance. This  suggests a mismatch between benchmarking of AI in isolation and in the context of human-AI teams.
\end{abstract}

\section{Introduction}
\label{sec:intro}

\begin{figure}[t!]
  \centering
  \includegraphics[width=0.78\linewidth]{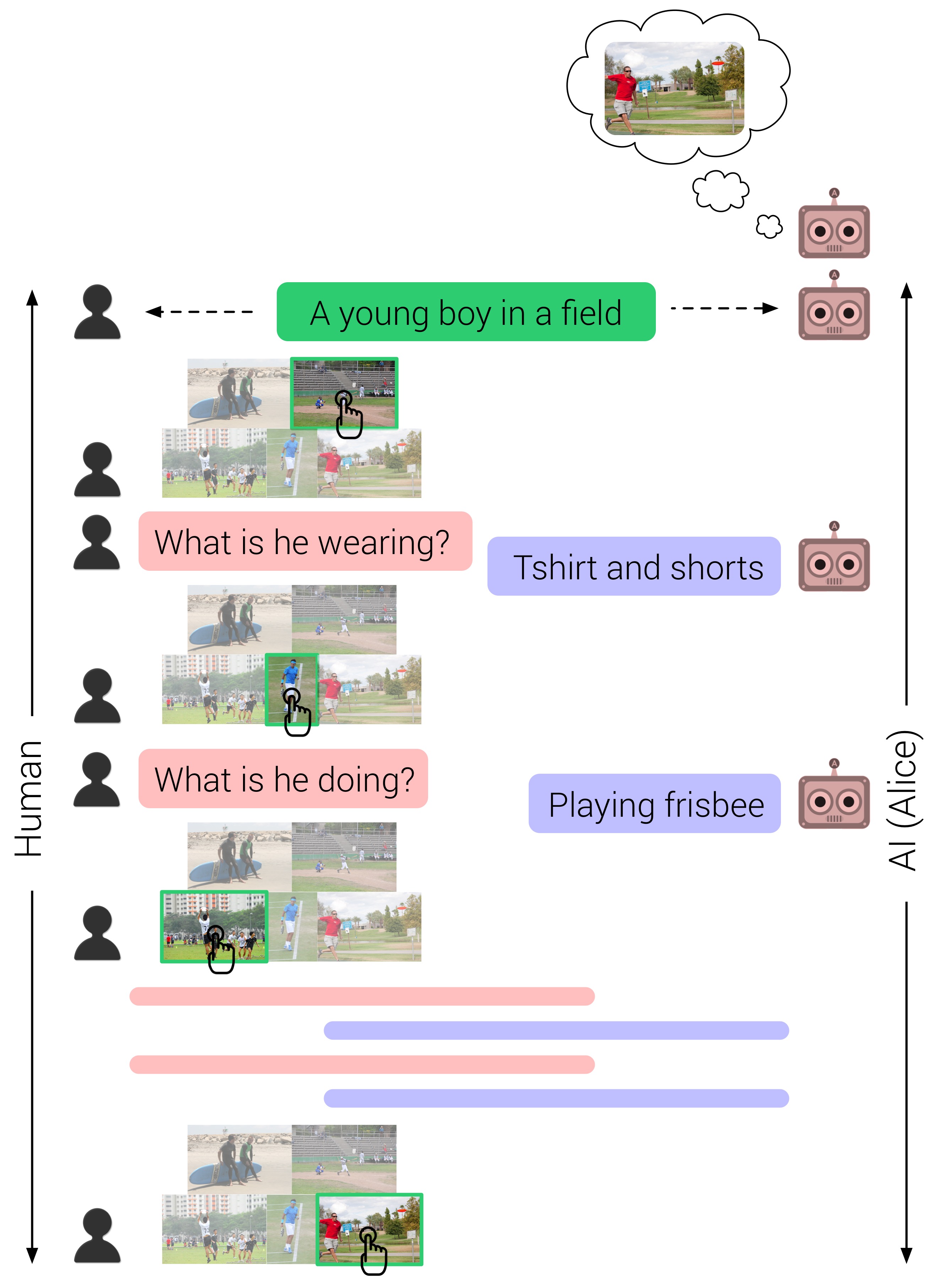}
  \caption{
A human and an AI (a visual conversation agent called \A) play the proposed \GW game.
At the start of the game (top), \A is provided an image (shown above \A) which is unknown to the human.
Both \A and the human are then provided a brief description of the image. The human then attempts to identify the secret image.
In each subsequent round of dialog, the human asks a question about the unknown image, receives an answer from \A, and makes a best guess of the secret image from a fixed pool of images.
After 9 rounds of dialog, the human makes consecutive guesses until the secret image is identified.
The fewer guesses the human needs to identify the secret image, the better the human-AI team performance.}
  \label{fig:teaser}
\end{figure}

As Artificial Intelligence (AI) systems become increasingly accurate and interactive (\eg Alexa, Siri, Cortana, Google Assistant), human-AI teams are inevitably going to become more commonplace.
To be an effective teammate, an AI must overcome the challenges involved with adapting to humans; however, progress in AI is routinely measured in isolation, without a human in the loop.
In this work, we focus specifically on the evaluation of visual conversational agents and develop a human computation game to benchmark their performance as members of human-AI teams.

Visual conversational agents~\cite{visdial,visdial_rl,vries_cvpr17,strub_arxiv17} are AI agents trained to understand and communicate about the contents of a scene in natural language.
For example, in \figref{fig:teaser}, the visual conversational agent (shown on the right) replies to answers questions about a scene while inferring context from the dialog history -- Human: "What is he doing?" Agent: "Playing frisbee".
These agents are typically trained to mimic large corpora of human-human dialogs and are evaluated automatically on how well they retrieve actual human responses (ground truth) in novel dialogs.

Recent work has evaluated these models more pragmatically by evaluating how well pairs of visual conversational agents perform on goal-based conversational tasks rather than response retrieval from fixed dialogs.
Specifically, \cite{visdial_rl} train two visual conversational agents -- a questioning bot \Qbot, and an answering bot \Abot~-- for an image-guessing task.
Starting from a description of the scene, \Qbot and \Abot converse over multiple rounds of questions (\Qbot) and answers (\Abot) in order to improve \Qbot's understanding of a secret image known only to \Abot.
After a fixed number of rounds, \Qbot must guess the secret image from a large pool and both \Qbot and \Abot are evaluated based on this guess.

\cite{visdial_rl} compare supervised baseline models with \Qbot-\Abot teams trained through reinforcement learning based self-talk on this image-guessing task. They find that the AI-AI teams improve significantly at guessing the correct image after self-talk updates compared to the supervised pretraining.
While these results indicate that the self-talk fine-tuned agents are better visual conversational agents, crucially, it remains unclear if these agents are indeed better at this task when \emph{interacting with humans}.

\xhdr{\GW.} In this work, we propose to evaluate if and how this progress in AI-AI evaluation translates to the performance of human-AI teams.
Inspired by the popular GuessWhat or 20-Questions game, we design a human computation game -- \GW~-- which requires collaboration between human and visual conversational AI agents.
Mirroring the setting of \cite{visdial_rl}, \GW is an image-guessing game that consists of 2 participants -- \emph{questioner} and \emph{answerer}.
At the start of the game, the answerer is provided an image that is unknown to the questioner and both questioner and answerer are given a brief description of the image content.
The questioner interacts with the answerer for a fixed number of rounds of question-answer (dialog) to identify the secret image from a fixed pool of images (see~\figref{fig:teaser}).

We evaluate human-AI team performance in \GW, for the setting where the questioner is a human and the answerer is an AI (that we denote \A). Specifically, we evaluate two versions of \A for \GW:
\begin{enumerate}
     \item \Asl which is trained in a supervised manner on the Visual Dialog dataset \cite{visdial} to mimic the answers given by humans when engaged in a conversation with other humans about an image, and
    \item \Arl which is pre-trained with supervised learning and fine-tuned via reinforcement learning for an image-guessing task as in \cite{visdial_rl}.
\end{enumerate}
It is important to appreciate the difficulty and sensitivity of the \GW game as an evaluation tool -- agents have to understand human questions and respond with accurate, consistent, fluent and informative answers for the human-AI team to do well.
Furthermore, they have to be robust to their own mistakes, \ie, if an agent makes an error at a particular round, that error is now part of its conversation history, and it must be able to correct itself rather than be consistently inaccurate. Similarly, human players must also learn to adapt to \A's sometime noisy and inaccurate responses.

At its core, \GW is a game-with-a-purpose (GWAP) that leverages human computation to evaluate visual conversational agents.
Traditionally, GWAP \cite{vonanh_acm2008} have focused on \emph{human-human collaboration}, \ie collecting data by making humans play games to label images \cite{vonahn_chi04}, music \cite{law2007tagatune} and movies \cite{michelucci_2013}.
We extend this to human-AI teams and to the best of our knowledge, our work is the first to evaluate visual conversational agents in an interactive setting where humans are continuously engaging with agents to succeed at a cooperative game.

\xhdr{Contributions.} More concretely, we make the following contributions in this work:
\begin{itemize}
    \item We design an interactive image-guessing game (\GW) for evaluating human-AI team performance in the specific context of the AIs being visual conversational agents.
    \GW pairs humans with \A, an AI capable of answering a sequence of questions about images. \A is assigned a secret image
    and answers questions asked about that image from a human for 9 rounds to help them identify the secret image (Sec.~\ref{sec:game}).
    \item We evaluate human-AI team performance
    on this game for both supervised learning (SL)
    and reinforcement learning (RL) versions of \A.
    Our main experimental finding is that despite significant differences between SL and RL agents reported in previous work~\cite{visdial_rl}, we find \emph{no significant difference} in performance between \Asl or \Arl when paired with human partners (Sec.~\ref{subsec:sl_vs_rl}).
This suggests that while self-talk and RL are interesting directions to pursue for building better visual conversational agents, there appears to be a  disconnect between AI-AI and human-AI evaluations -- progress on former does not seem predictive of progress on latter. This is an important finding to guide future research.
\end{itemize}

\section{Related Work}

Given that our goal is to evaluate visual conversational agents through a human computation game, we draw connections to relevant work on visual conversational agents, human computation games, and dialog evaluation below.

\xhdr{Visual Conversational Agents.}
Our AI agents are visual conversational models, which have recently emerged as a popular research area
in visually-grounded language modeling~\cite{visdial,visdial_rl,vries_cvpr17,strub_arxiv17}.
\cite{visdial} introduced the task of Visual Dialog and collected the VisDial dataset by pairing
subjects on Amazon Mechanical Turk (AMT) to chat about an image (with assigned roles of questioner and answerer).
\cite{visdial_rl} pre-trained questioner and answerer agents on this VisDial dataset via supervised
learning and fine-tuned them via self-talk (reinforcement learning), observing that RL-fine-tuned \Qbot-\Abot
are better at image-guessing after interacting with each other.
However, as described in \secref{sec:intro}, they do not evaluate if this change in \Qbot-\Abot performance translates to human-AI teams.

\xhdr{Human Computation Games.}
Human computation games have been shown to be time- and cost-efficient, reliable,
intrinsically engaging for participants \cite{jain_acm2013,krause2011human}, and
hence an effective method to collect data annotations.
There is a long line of work on designing such Games with a Purpose (GWAP) \cite{vonanh_acm2008} for data labeling
purposes across various domains including images \cite{vonahn_chi04,von2006peekaboom,law2009input,kazemzadeh_emnlp14},
audio \cite{diakopoulos2008audio,law2007tagatune}, language \cite{aras2010webpardy,chamberlain2008phrase},
movies \cite{michelucci_2013} \etc. While such games have traditionally focused on human-human collaboration,
we extend these ideas to human-AI teams. Rather than collecting labeled data,
our game is designed to measure the effectiveness of the AI in the context of human-AI teams.

\xhdr{Evaluating Conversational Agents.}
Goal-driven (non-visual) conversational models have typically been evaluated on task-completion rate or time-to-task-completion \cite{paek_elds01}, so shorter conversations are better.
At the other end of the spectrum, free-form conversation models are often evaluated by metrics that rely on n-gram overlaps, such as BLEU, METEOR, ROUGE, but these have been shown to correlate poorly with human judgment \cite{liu_emnlp16}.
Human evaluation of conversations is typically in the format where humans rate the quality of machine utterances given context, without actually taking part in the conversation, as in \cite{visdial_rl} and \cite{li_emnlp16}.
To the best of our knowledge, we are the first to evaluate conversational models via team performance where humans are continuously interacting with agents to succeed at a downstream task.

\xhdr{Turing Test.}
Finally, our \GW game is in line with ideas in \cite{grosz2012question}, re-imagining the traditional Turing Test
for state-of-the-art AI systems, taking the pragmatic view that an effective AI teammate need not appear human-like, act or be mistaken for one, provided its behavior does not feel jarring or baffle teammates, leaving them wondering not about what it is thinking but whether it is.

Next, we formally define the AI agent \A (Sec.~\ref{sec:AI}), describe the \GW game setup (Sec.~\ref{sec:game} and \ref{sec:infra}), and present results and analysis from human studies (Sec.~\ref{sec:results}).

\begin{figure*}
    \centering
    \includegraphics[width=0.85\linewidth]{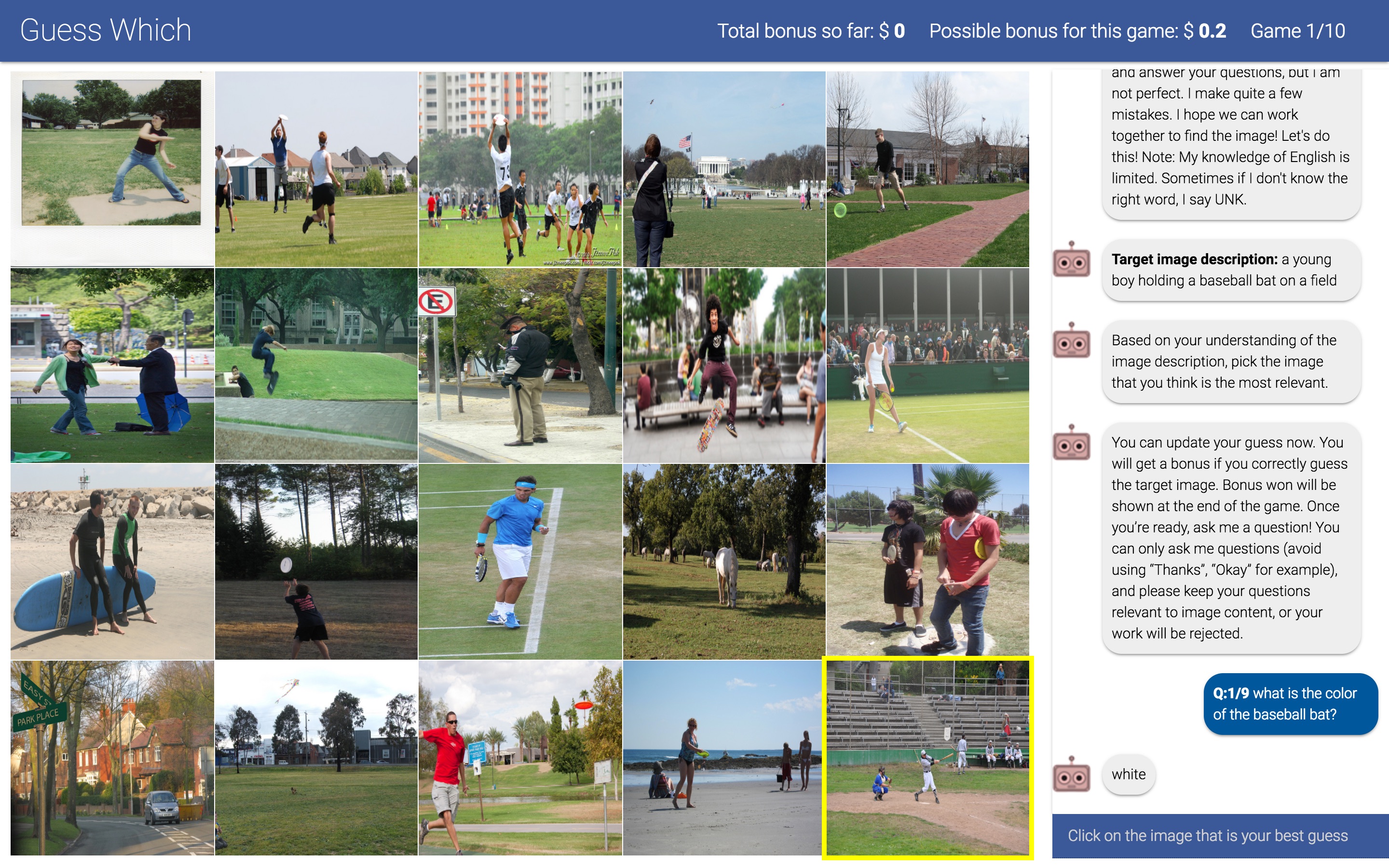}
    \caption{\GW Interface: A user asks a question to \A in each round and \A responds with an answer. The user then selects an appropriate image which they think is the secret image after each round of conversation. At the end of the dialog, user successively clicks on their best guesses until they correctly identify the secret image.}
    \label{fig:gw_interface}
\end{figure*}

\section{The AI: \A}
\label{sec:AI}

Recall from \secref{sec:intro} that our goal is to evaluate how progress in AI measured through automatic evaluation translates to performance of human-AI teams in the context of visual conversational agents.
Specifically, we are considering the question-answering agent \Abot from \cite{visdial_rl} as \Abot is the agent more likely to be deployed with a human partner in real applications (\eg to answer questions about visual content to aid a visually impaired user).
For completeness, we will review this work in this section.

\cite{visdial_rl} formulate a self-supervised image-guessing task between a questioner bot (\Qbot) and an answerer bot (\Abot) which plays out over multiple rounds of dialog.  At the start of the task, \Qbot and \Abot are shown a one sentence description (\ie a caption) of an image (unknown to \Qbot). The pair can then engage in question and answer based dialog for a fixed number of iterations after which \Qbot must try to select the secret image from a pool. The goal of the \Qbot-\Abot team is two-fold, \Qbot should: 1) build a mental model of the unseen image purely from the dialog and 2) be able to retrieve that image from a line-up of images.

Both \Qbot and \Abot are modeled as Hierarchical Recurrent Encoder-Decoder neural networks \cite{visdial,serban_aaai16} which encode each round of dialog independently via a recurrent neural network (RNN) before accumulating this information through time with an additional RNN (resulting in hierarchical encoding). This representation (and a convolutional neural network based image encoding in \Abot's case) are used as input to a decoder RNN which produces an agent's utterance (question for \Qbot and answer for \Abot) based on the dialog (and image for \Abot). In addition, \Qbot includes an image feature regression network that predicts a representation of the secret image based on dialog history. We refer to \cite{visdial_rl} for complete model details.

These agents are pre-trained with supervised dialog data from the VisDial dataset ~\cite{visdial} with a Maximum Likelihood Estimation objective. This pre-training ensures that agents can generally recognize objects/scenes and utter English. Following this, the models are fine-tuned by `smoothly' transitioning to a deep reinforcement learning framework to directly improve image-guessing performance. This annealed transition avoids abrupt divergence of the dialog in face of an incorrect question-answer pair in the \Qbot-\Abot exchange.
During RL based self-talk, the agents' parameters are updated by gradients corresponding to rewards depending on individual good or bad exchanges.
We refer to the baseline supervised learning based \Abot as \Asl and the RL fine-tuned bot as \Arl. \cite{visdial_rl} found that the AI-AI pair succeeds in retrieving the correct image more often after being fine-tuned with RL.
In the following section, we outline our \GW game designed to evaluate whether this improvement between \Asl and \Arl in automatic metrics translates to human-AI collaborations.

\section{Our \GW Game}
\label{sec:game}

We begin by describing our game setting; outlining the players and gameplay mechanics. A video of an example game being played can be found at {\tt{\small \url{https://vimeo.com/229488160}}}.

\xhdr{Players.} We replace \Qbot in the AI-AI dialog with humans to perform a collaborative task of identifying a secret image from a pool. In the following, we will refer to \Abot as \A and the human player as \hum. We evaluate two versions of \A~-- \Asl and \Arl, where SL and RL correspond to agents \emph{trained in a supervised setting} and \emph{fine-tuned with reinforcement learning} respectively.

\xhdr{Gameplay.}
In our game setting, \A is assigned a secret image $I^{c}$ (unknown to \hum) from a pool of images $\mathbb{I} = \{I_{1},I_{2},...,I_{n}\}$ taken from the COCO dataset~\cite{mscoco}.
Prior to beginning the dialog, both \A and \hum are provided a brief description (\ie a caption) of $I^{c}$ generated by Neuraltalk2~\cite{ntk2}, an open-source implementation of~\cite{vinyals_cvpr15}.
\hum then makes a guess about the secret image by selecting one from the pool $\mathbb{I}$ based only on the caption, \ie before the dialog begins.

In each of the following rounds, \hum asks \A a question $q_t$ about the secret image $I^{c}$ in order to better identify it from the pool and \A responds with an answer $a_t$. After each round, \hum must select an image $I^{t}$ that they feel is most likely the secret image $I^{c}$ from pool $\mathbb{I}$ based on the dialog so far. At the end of $k=9$ rounds of dialog, \hum is asked to successively click on their best guess. At each click, the interface gives \hum feedback on whether their guess is correct or not and this continues until \hum guesses the true secret image. In this way, \hum induces a partial ranking of the pool up to the  secret image based on their mental model of $I^{c}$ from the dialog.

\subsection{Pool Selection}
When creating a pool of images, our aim is to ensure that the game is challenging and engaging, and not too easy or too hard. Thus, we construct each pool of images $\mathbb{I}$ in two steps -- first, we choose the secret image $I^{c}$, and then sample similar images as distractors for $I^{c}$. \figref{fig:gw_interface} shows a screenshot of our game interface including a sample image pool and chat.

\xhdr{Secret Image Selection.} VisDial v0.5 is constructed on $68k$ COCO images which contain complex everyday scenes with 80 object categories. \Abot
is trained and validated on VisDial v0.5 \emph{train} and \emph{val} splits respectively. As the images for both these splits come from COCO-train, we sample secret images and pools from COCO-validation to avoid overlap.

To select representative secret images and diverse image pools, we do the following. For each image in the COCO validation set, we extract the penultimate layer (`fc7') activations of a standard deep convolutational neural network (VGG-19 from~\cite{simonyan_iclr15}). For each of the 80 categories, we average the embedding vector of all images containing that category. We then pick those images closest to the mean embeddings, yielding 80 candidates.

\xhdr{Generating Distractor Images.} The distractor images are designed to be semantically similar to the secret image $I^{c}$.
For each candidate secret image, we created 3 concentric hyper-spheres as euclidean balls (of radii increasing in arithmetic progression) centered on the candidate secret image in fc7 embedding space, and sampled images from each sphere in a fixed proportion to generate a pool corresponding to the secret image.
The radius of the largest sphere was varied and manually validated to ensure pool difficulty.
The sampling proportion can be varied to generate pools of varying difficulty. Of the 80 candidate pools, we picked 10 that were of medium difficulty based on manual inspection.

\subsection{Data Collection and Player Reward Structure}
We use AMT to solicit human players for our game. Each Human Intelligence Task (HIT) consists of 10 games (each game corresponds to one pool) and we find that overall 76.7\% of users who started a HIT completed it \ie played all 10 games. We note that incomplete game data was discarded and does not contribute to the analysis presented in subsequent sections.

We published HITs until 28 games with both \Asl and \Arl were completed. This results in a total of 560 games split between the agents, with each game consisting of 9 rounds of dialog and 10 rounds of guessing.
Workers are paid a base pay of \$5 per HIT ($\sim$\$10/hour).

To incentivize workers to try their best at guessing the secret image,
    workers are paid a two-part bonus --
  (1) based on the number of times their best guess matched the true secret image after each round (up to \$1 per HIT), and
    (2) based on the rank of the true secret image in their final sorting at the end of dialog (up to \$2 per HIT).

This final ranking explicitly captures the workers' mental model of the secret image
(unlike the per-round, best-guess estimates),
and is closer to the overall purpose of the game (identifying the secret image at the end of the dialog). As such, this final sorting is given a higher potential bonus.

\subsection{Evaluation}
\label{sec:eval}
Since the game is structured as a retrieval task, we evaluate the human-AI collaborative performance using standard retrieval metrics. Note that the successive selection of images by \hum at the end of the dialog tells us the rank of the true secret image in a sorting of the image pool based on \hum's mental model. For example, if \hum makes 4 guesses before correctly selecting the secret image, then \hum's mental model ranked the secret image 5th within the pool.

To evaluate human-AI collaboration, we use the following metrics: (1) Mean Rank (MR), which is the mean rank of the secret image (\ie number of guesses it takes to identify the secret image). Lower values indicate better performance. (2) Mean Reciprocal Rank (MRR), which is the mean of the reciprocal of the rank of the secret image. MRR penalizes differences in lower ranks (e.g., between 1 and 2) greater than those in higher ranks (e.g., between 19 and 20). Higher values indicate better performance.

\begin{figure*}[t]
\centering
\setlength{\fboxsep}{0pt}
\setlength{\fboxrule}{0pt}
    \begin{subfigure}[t]{3.2in}
    \fbox{\includegraphics[width=3.2in,height=1.8in]{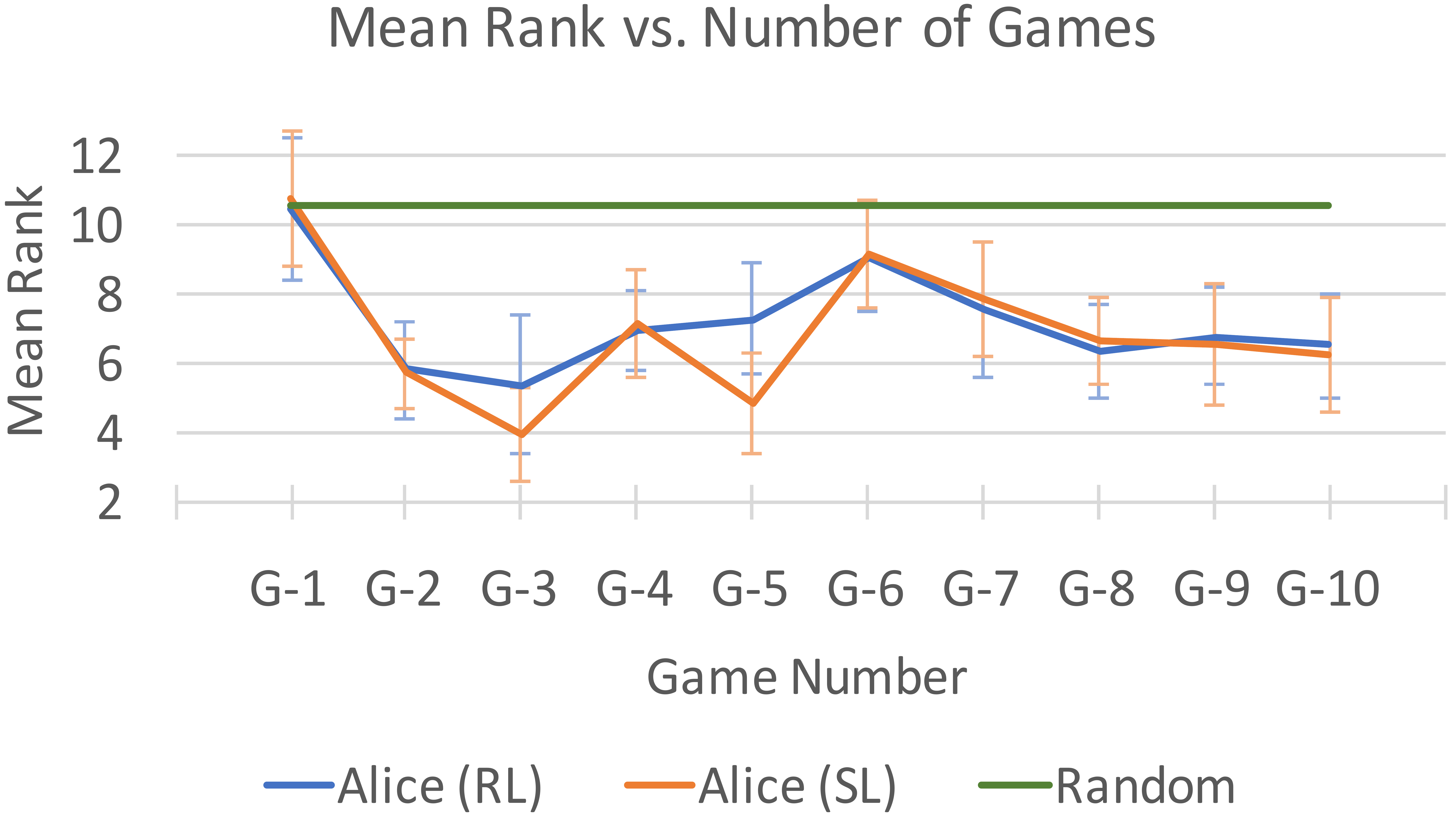}}
    \caption{\Asl and \Arl perform about the same for most games and outperform a baseline model that makes a string of random guesses at the end of each game.}
    \label{fig:games_time}
    \end{subfigure}
    \hspace{0.4in}
    \begin{subfigure}[t]{3.2in}
    \fbox{\includegraphics[width=3.2in,height=1.8in]{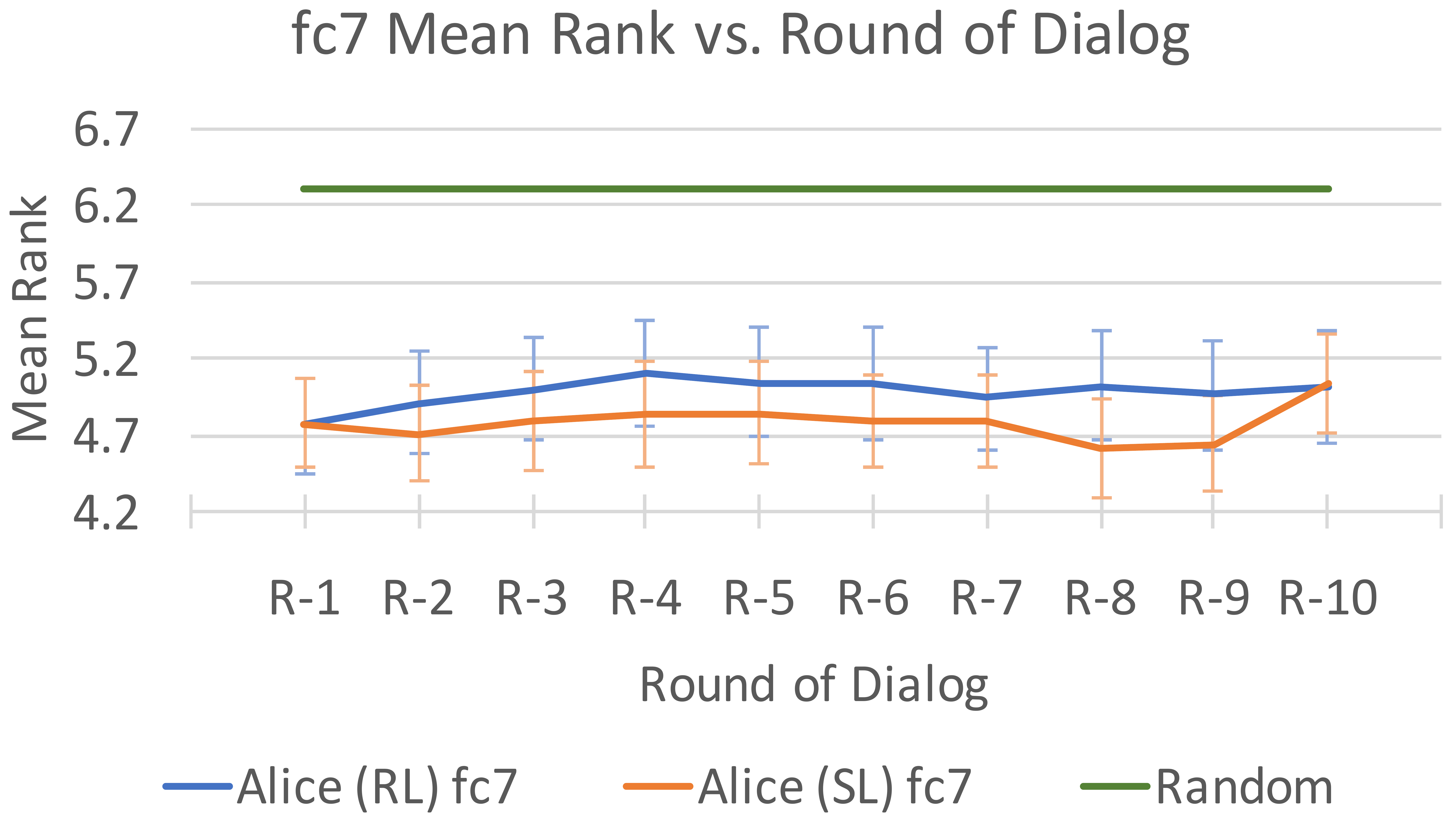}}
    \caption{\Asl and \Arl perform about the same, and clearly outperform a baseline model that randomly chooses an image. As described in Sec.~\ref{sec:eval}, this is only a coarse estimate of the rank of the secret image after each round of dialog.}
  \label{fig:rounds}
  \end{subfigure}
  \caption{Mean rank (MR) of secret image across (a) number of games and (b) rounds of dialog. Lower is better. Error bars are 95\% confidence intervals from 1000 bootstrap samples.}
\end{figure*}

At the end of each round, \hum makes their best guess of the secret image. To get a coarse estimate of the rank of the secret image in each round, we sort the image pool based on distance in fc7 embedding space from \hum's best guess.
This can be used to assess accuracy of \hum's mental model of the secret image after each round of dialog (e.g., Fig.~\ref{fig:rounds}).

\begin{figure}
    \centering
    \includegraphics[width=0.8\linewidth]{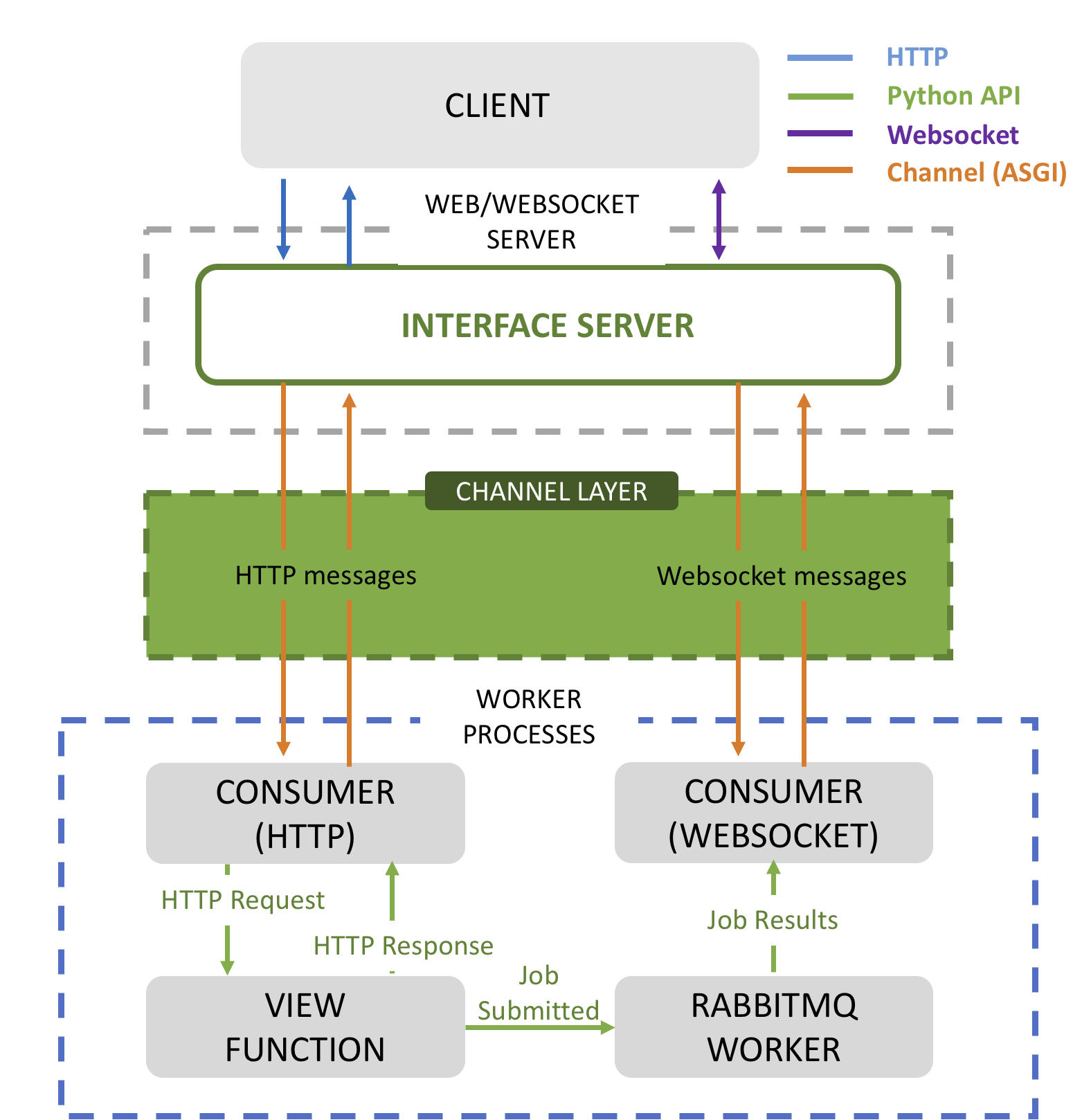}
    \caption{We outline the backend architecture of our implementation of \gw. Since \gw requires a live interaction between the human and the AI, we design a workflow that can handle multiple queues and can quickly pair a human with an AI agent.}
    \label{fig:gw_pipeline}
\end{figure}

\section{Infrastructure}
\label{sec:infra}
We briefly outline the backend architecture of \GW in this section. Unlike most human-labeling tasks that are one-way and static in nature (i.e., only involving a human labeling static data), evaluating AI agents via our game requires live interaction between the AI agent and the human.
We develop a robust workflow that can maintain a queue of workers and pair them up in real-time with an AI agent.

We deploy \Asl and \Arl on an AWS EC2~\cite{aws} GPU instance.
We use Django (a Model-View-Controller web framework written in Python) which helps in monitoring HITs in real-time.
We use~\cite{rabbitmq}, an open source message broker, to queue inference jobs that generate dialog responses from the model.
Our backend is asynchronously connected to the client browser via websockets such that
whenever an inference job is completed, a
websocket polls the AI response and delivers it to the human in real-time.
We store and fetch data efficiently to and from a PostgreSQL database.
\reffig{fig:gw_pipeline} shows a schematic diagram of the backend architecture.
Our complete backend infrastructure and code will be made publicly available for others to easily make use of our human-AI game interface.

\section{Results}
\label{sec:results}

\subsection{\Asl vs. \Arl}
\label{subsec:sl_vs_rl}

We compare the performance of the two agents \Asl and \Arl in the \gw game. These bots are state-of-the-art visual dialog agents with respect to emulating human responses and generating visually discriminative responses in AI-AI dialog.~\cite{visdial_rl} evaluate these agents against strong baselines and report AI-AI team results that are significantly better than chance on a pool of $\sim$10k images (rank $\sim$1000 for SL, rank $\sim$500 for RL). In addition to evaluating them in the context of human-AI teams we also report \Qbot-\A team performances for reference.

In Table~\ref{tab:asl_vs_arl}, we compare the performances of human-\Asl and human-\Arl teams according to Mean Rank (MR) and Mean Reciprocal Rank (MRR) of the secret image based on the guesses \hum makes at the end of dialog.
We observe that at the end of each game (9 rounds of dialog), human subjects correctly guessed the secret image on their 6.86th attempt (Mean Rank) when \Asl was their teammate. With \Arl as their teammate, the average number of guesses required was 7.19. We also observe that \Arl outperforms \Asl on the MRR metric.
On both metrics, however, the differences are within
the standard error margins (reported in the table) and not statistically significant. As we collected additional data, the error margins became smaller but the means also became closer. This interesting finding stands in stark contrast to the results reported by~\cite{visdial_rl},
where
\Arl was found to be significantly more accurate than \Asl
when evaluated in an AI-AI team.
Our results suggest that the improvements
of RL over SL (in AI-AI teams)
do not seem to translate to
when the agents are paired with a human in a similar setting.

\begin{table}[t]
\setlength{\tabcolsep}{8pt}
{\small
\begin{center}
{
\begin{tabular}{lcc@{}}
\toprule
\textbf{Team} & \textbf{MR} & \textbf{MRR} \\
\toprule
Human-\Asl & 6.86 $\pm$ 0.53 & \multicolumn{1}{r}{0.27 $\pm$ 0.03} \\
\midrule
Human-\Arl & 7.19 $\pm$ 0.55 & \multicolumn{1}{r}{0.25 $\pm$ 0.03} \\
\bottomrule
\end{tabular}}
\end{center}
\caption{Performance of Human-\A teams with \Asl and \Arl measured by MR (lower is better) and MRR (higher is better). Error bars are 95\% CIs from 1000 bootstrap samples. Unlike (Das et al., 2017b), we find no significant difference between \Asl and \Arl.}
\label{tab:asl_vs_arl}
}
\end{table}

\xhdr{MR with varying number of games.} In Fig.~\ref{fig:games_time}, we plot the mean rank (MR) of the secret image across different games. We see that the human-\A team performs about the same for both \Asl and \Arl except Game 5, where \Asl seems to marginally outperform \Arl. We compare the performance of these teams against a baseline model that makes a string of random guesses at the end of the game. The human-\A teams outperforms this random baseline with a relative improvement of about 25\%.

\begin{table}[t]
\setlength{\tabcolsep}{8pt}
{\small
\begin{center}
{
\begin{tabular}{lcc@{}}
\toprule
\textbf{Team} & \textbf{\Asl} & \textbf{\Arl} \\
\toprule
Human & 6.9 & \multicolumn{1}{r}{7.2} \\
\midrule
\Qbot (SL) & 5.6 & \multicolumn{1}{r}{5.3} \\
\midrule
\Qbot (RL) & 4.7 & \multicolumn{1}{r}{4.7} \\
\bottomrule
\end{tabular}}
\end{center}
\caption{Performance of Human-\A and \Qbot-\A teams measured by MR (lower is better). We observe that AI-AI teams outperform human-AI teams.}
\label{tab:mr_matrix}}
\end{table}

\xhdr{AI-\A teams versus human-\A teams.} In Table~\ref{tab:mr_matrix}, we compare team performances by pairing three kinds of questioners -- human, \Qbot (SL) and \Qbot (RL) with \Asl and \Arl (6 teams in total) to gain insights about how the questioner and \A influence team performances. Interestingly, we observe that AI-\A teams outperform human-\A teams. On average, a \Qbot (SL)-\Asl team takes about 5.6 guesses to arrive at the correct secret image (as opposed to 6.86 guesses for a human-\Asl team). Similarly, a \Qbot (RL)-\Arl team takes 4.7 guesses as opposed to a human-\Arl team which takes 7.19 guesses. When we compare AI-AI teams (see Row 2 and 3) under different settings, we observe that teams having \Qbot (RL) as the questioner outperform those with \Qbot (SL). Qualitatively, we found that \Qbot (SL) tends to ask repeating questions in a dialog and that questions from \Qbot (RL) tend to be more visually grounded compared to \Qbot (SL). Also, note that among the four teams \A does not seem to affect performance across SL and RL.

Since we observe that \Qbot (RL) tends to be a better questioner on average compared to \Qbot (SL), as future work, it will be interesting to explore a setting where we evaluate \Qbot via a similar game with the human playing the role of answerer in a \Qbot-human team.

\xhdr{MR with varying rounds of dialog.} \figref{fig:rounds} shows a coarse estimate of the mean rank of the secret image across rounds of a dialog, averaged across games and workers. As explained in Sec.~\ref{sec:eval}, image ranks are computed via distance in embedding space from the guessed image (and hence, are only an estimate). We see that the human-\A team performs about the same for both \Asl and \Arl across rounds of dialog in a game. When compared with a baseline agent that makes random guesses after every round of dialog, the human-\A team clearly performs better.\\\\

\xhdr{Statistical tests.}  Observe that on both the metrics (MR and MRR), the differences between performances of \Asl and \Arl are within error margins. Since both standard error and bootstrap based $95\%$ confidence intervals overlap significantly, we ran further statistical tests. We find no significant difference between the mean ranks of \Asl and \Arl under a Mann-Whitney U test ($p=0.44$).

\begin{figure}[t]
    \includegraphics[scale=0.23]{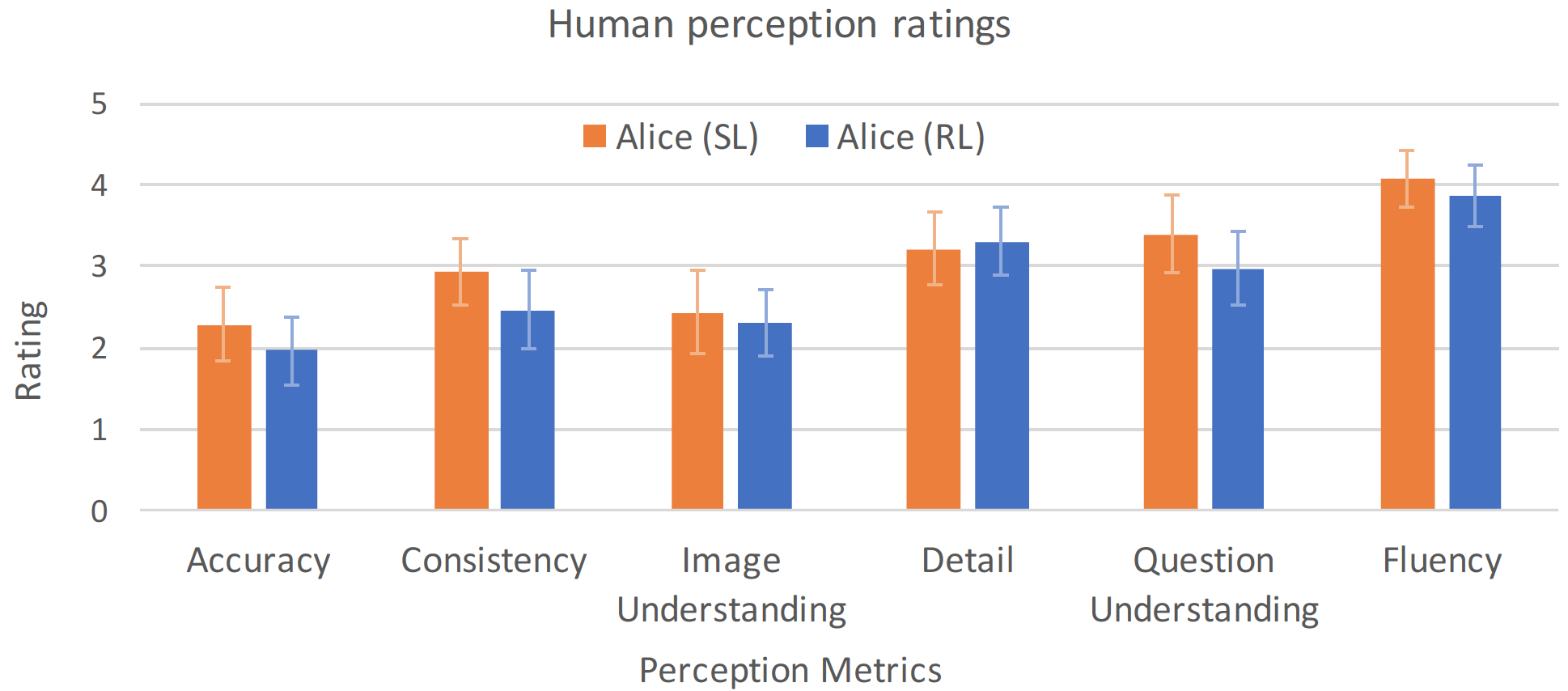}
    \caption{Worker ratings for \Asl and \Arl on 6 metrics. Higher is better. Error bars are 95\% confidence intervals from 1000 bootstrap samples. Humans perceive no significant differences between \Asl and \Arl across the 6 feedback metrics.}
    \label{fig:feedback}
\end{figure}

\subsection{Human perception of AI teammate}

At the end of each HIT, we asked workers for feedback on \A. Specifically, we asked workers to rate \A on a 5-point scale (where 1=Strongly disagree, 5=Strongly agree), along 6 dimensions. As shown in~\figref{fig:feedback}, \A was rated on -- how accurate they thought it was (accuracy), how consistent its answers were with its previous answers (consistency), how well it understood the secret image (image understanding), how detailed its answers were (detail), how well it seemed to understand their questions (question understanding) and how fluent its answers were (fluency).

We see in~\figref{fig:feedback} that humans perceive both \Asl and \Arl as comparable in terms of all metrics. The small differences in perception are not statistically significant.

\subsection{Questioning Strategies}
\begin{figure}[t]
    \centering
    \includegraphics[scale=0.18]{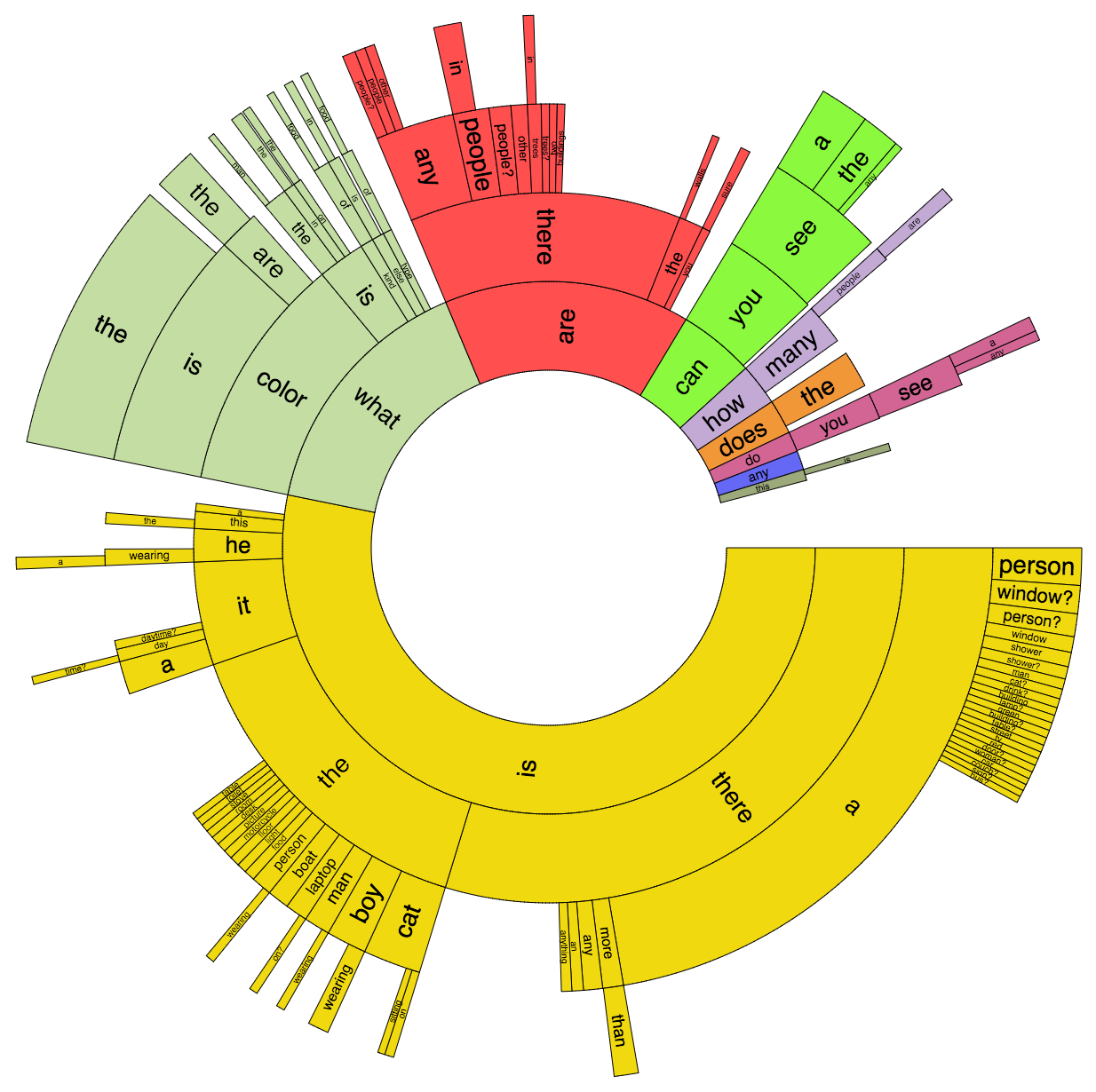}
    \caption{Distribution of first n-grams for questions asked to \A. Word ordering starts from the center and radiates outwards. Arc length is proportional to the number of questions containing the word. The most common question-types are binary -- followed by `What color..' questions.}
    \vspace{-10pt}
    \label{fig:ques_circles}
\end{figure}

\begin{figure*}
    \centering
    \includegraphics[width=\textwidth]{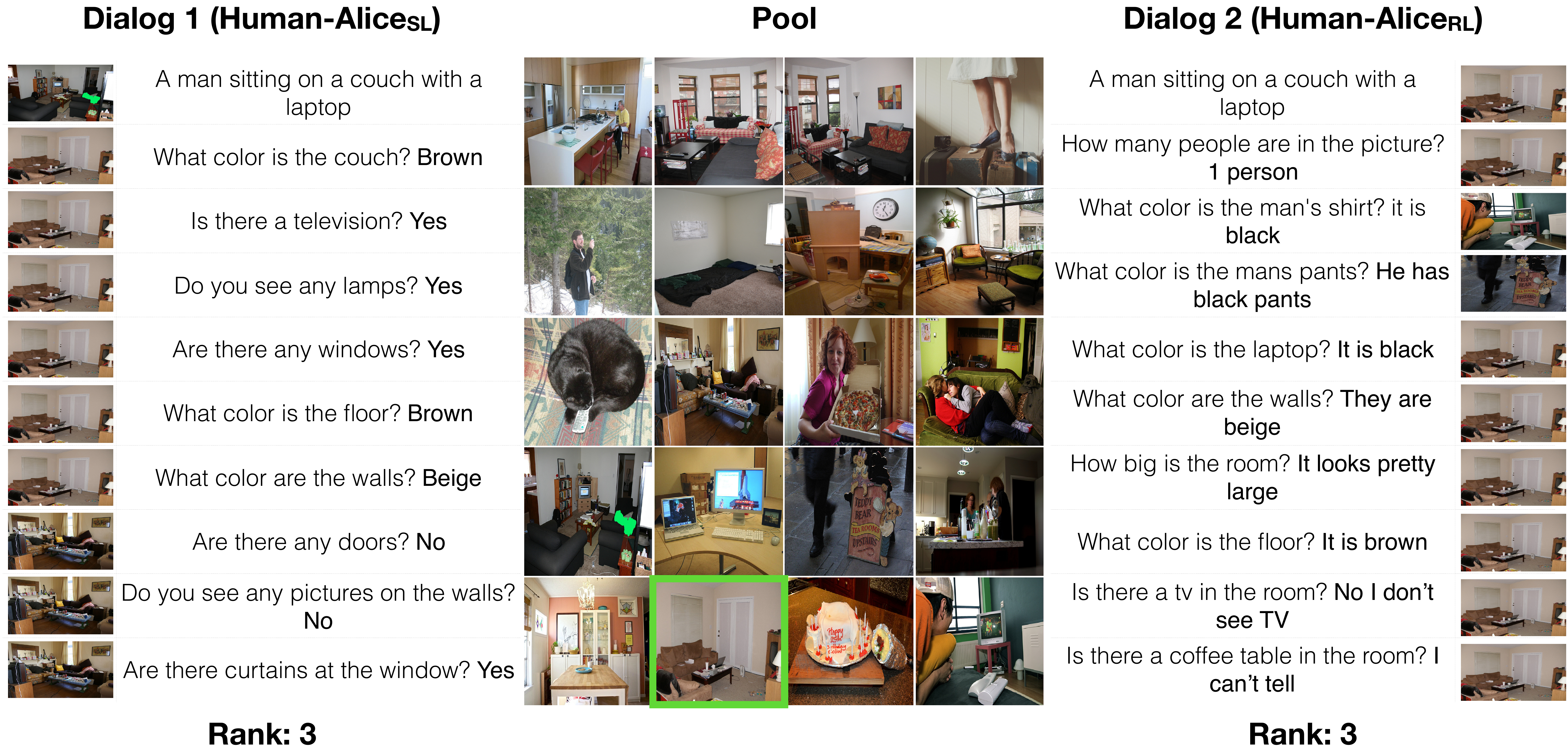}
    \caption{We contrast two games played by different workers with \Asl and \Arl on the same pool (secret image outlined in green). In both cases, the workers are able to find the secret image within three guesses. It is also interesting to note how the answers provided by \A are different in the two cases.}
    \label{fig:qualitative}
    \vspace{-8pt}
\end{figure*}

\figref{fig:ques_circles} shows the distribution of questions that human subjects ask \A in \GW. Akin to the format of the human-human GuessWhat game, we observe that binary (yes/no) questions are overwhelmingly the most common question type, for instance, ``Is there/the/he ...?'' (region shaded yellow in the figure), ``Are there ...?'' (region shaded red), etc. The next most frequent question is ``What color ...?''. These questions may be those that help the human discriminate the secret image the best. It could also be that humans are attempting to play to the perceived strengths of \A. As people play multiple games with \A, it is possible that they discover \A's strengths and learn to ask questions that play to its strengths. Another common question type is counting questions, such as ``How many ...?''. Interestingly, some workers adopt the strategy of querying \A with a single word (e.g., nouns such as ``people'', ``pictures'', etc.) or a phrase (e.g., ``no people'', ``any cars'', etc.). This strategy, while minimizing human effort, does not appear to change \A's performance.
~\figref{fig:qualitative} shows a game played by two different subjects.

\section{Challenges}
\par \noindent
There exist several challenges that are unique to human computation in the context of evaluating human-AI teams, for instance, making our games engaging while still ensuring fair and accurate evaluation. In this section, we briefly discuss some of the challenges we faced and our solutions to them.
\xhdr{Knowledge Leak.} It has been shown that work division in crowdsourcing tasks follows a Pareto principle~\cite{little2009many}, as a small fraction of workers usually complete
 a majority of the work. In the context of evaluating an AI based on performance of a human-AI team, this poses a challenge.

Recently,~\cite{toaim} showed that human subjects can predict the responses of an AI more accurately with higher familiarity with the AI.
That is, a human's knowledge gained from familiarity with their AI teammate, can bias the performance of the human-AI team -- knowledge from previous tasks might leak to later tasks.
To prevent a biased evaluation of team performance due to human subjects who have differing familiarity with \A, every person only plays a fixed number of games (10) with \A. Thus, a human subject can only accept one task on AMT, which involves playing 10 games. The downside to this is that our ability to conduct a fair evaluation of an AI in an interactive, game-like setting is constrained by the number of unique workers who accept our tasks.

\xhdr{Engagement vs. Fairness.} In order to improve user-engagement while playing our games, we offer subjects performance-based incentives that are tied to the success of the human-AI team. There is one potential issue with this however. Owing to the inherent complexity of the visual dialog task, \A tends to be inaccurate at times. This increases both the difficulty and unpredictability of the game, as it tends to be more accurate for certain types of questions compared to others. We observe that this often leads to unsuccessful game-plays, sometimes due to errors accumulating from successive incorrect responses from \A to questions from the human. In a few other cases, the human is misled by \A by a single wrong answer or by the seed caption that tends to be inaccurate at times. While we would like to keep subjects engaged in the game to the best extent possible by providing performance-based incentives, issuing a performance bonus that depends on both the human and \A (who is imperfect), can be dissatisfying. To be fair to the subjects performing the task while still rewarding good performance, we split our overall budget for each HIT into a suitable fraction between the base pay (majority), and the performance bonus.

\section{Conclusion}

In contrast to the common practice of measuring AI progress in isolation, our work proposes benchmarking AI agents via interactive downstream tasks (cooperative games) performed by human-AI teams. In particular, we evaluate visual conversational agents in the context of human-AI teams. We design a cooperative game -- \GW~-- that involves a human engaging in a dialog with an answerer-bot (\A) to identify a secret image known to \A but unknown to the human from a pool of images. At the end of the dialog, the human is asked to pick out the secret image from the image pool by making successive guesses. We find that \Arl (fine-tuned with reinforcement learning) that has been found to be more accurate in AI literature than it's supervised learning counterpart when evaluated via a questioner bot (\Qbot)-\A team, is not more accurate when evaluated via a human-\A team. This suggests that there is a disconnect between between benchmarking of AI in isolation versus in the context of human-AI interaction. An interesting direction of future work could be to evaluate \Qbot via \Qbot-human teams.

We describe the game structure and the backend architecture and discuss the unique computation and infrastructure challenges that arise when designing such live interactive settings on AMT relative to static human-labeling tasks. Our code and infrastructure will be made publicly available.

\xhdr{Acknowledgements.}  We would like to acknowledge the effort provided by workers on Amazon Mechanical Turk. We are grateful to the developers of Torch~\cite{torch} for building an excellent framework. This work was funded in part by NSF CAREER awards to DB and DP, ONR YIP awards to DP and DB, ONR Grant N00014-14-1-0679 to DB, ONR Grant N00014-16-1-2713 to DP, a Sloan Fellowship to DP, an Allen Distinguished Investigator award to DP from the Paul G. Allen Family Foundation, Google Faculty Research Awards to DP and DB, Amazon Academic Research Awards to DP and DB, AWS in Education Research grant to DB, and NVIDIA GPU donations to DB. SL was partially supported by the Bradley Postdoctoral Fellowship. The views and conclusions contained herein are those of the authors and should not be interpreted as necessarily representing the official policies or endorsements, either expressed or implied, of the U.S. Government, or any sponsor.

{
\small
\bibliography{strings,main}
\bibliographystyle{aaai}
}

\end{document}